# Electric field control of third-order nonlinear Hall effect

Jiaju Yang[1, #], Lujun Wei[1, 2, #, *], Yanghui Li[1], Lina Chen[1], Wei Niu[1], Jiarui Chen[2], Jun Du[3, 4, *], and Yong Pu[1, *]

[1]*School of Science, Nanjing University of Posts and Telecommunications, Nanjing 210023, P. R. China*

[2]*National Laboratory of Solid State Microstructures, Nanjing University, Nanjing 210093, China*

[3]*Department of Physics, Nanjing University, Nanjing 210093, China*

[4]*National Key Laboratory of Spintronics, Nanjing University, Suzhou 215163, China*

[#] Jiaju Yang and Lujun Wei contributed equally to this work.

*Corresponding authors: wlj@njupt.edu.cn; jdu@nju.edu.cn; and yongpu@njupt.edu.cn.

## ABSTRACT

**The third-order nonlinear Hall effect (NLHE) serves as a sensitive probe of energy band geometric property, providing a new paradigm for revealing the Berry curvature distribution and topological response of quantum materials. In the type-_II_ Weyl semimetal TaIrTe$_4$, we report for the first time that the sign of the third-order NLHE reverses with decreasing temperature. Through scaling law analysis, we think that the third-order NLHE at high (T > 23 K) and low (T < 23 K) temperatures is dominated by Berry-connection polarizability (BCP) and impurity scattering, respectively. The third-order NLHE response strength can be effectively modulated by an additional applied in-plane constant electric field. At the high temperature region, the BCP reduction induced by the electric field leads to a decrease in the third-order NLHE response strength, while at the low temperature region, the electric field cause both BCP and impurity scattering**

effects to weaken, resulting in a more significant modulation of the third-order NLHE response strength. At 4 K and an electric field strength of 0.3 kV/cm, the modulated relative response strength could reach up to 65.3%. This work provides a new means to explore the third-order NLHE and a valuable reference for the development of novel electronic devices.

**Keywords:** Weyl semimetal TaIrTe$_4$, third-nonlinear Hall effect, electric-field control of third-nonlinear Hall effect, Berry-connection polarizability, impurity scattering effect

**Introduction**

The third-order nonlinear Hall effect (NLHE), as a novel quantum transport phenomenon, is characterized by a third-order transverse voltage response generated under longitudinal current excitation. Unlike the conventional Hall effect, this effect does not require time-reversal symmetry breaking, and its physical nature is determined by the geometrical properties of the energy bands characterized by the Berry-connection polarizability (BCP) tensor [1, 2], which is correlated with the symmetry of the material structure. Since the first experimental observation realized in $T_d$-phase MoTe$_2$ [1], third-order NLHE has been observed in WTe$_2$ (space group $Pmn2_1$) [3-5]、Cd$_3$As$_2$ (antispinel structure) [6]、1$T$-VSe$_2$ (hexagonal layered structure and the trigonal space group $P$-$3m1$)1$T$-VSe$_2$ [7], monolayer graphene [8] and MnBi$_2$Te$_4$ [9] have been successively demonstrated in topological material systems.

Notably, the type-$II$ semimetallic TaIrTe$_4$ exhibits second- and third-order NLHEs at room temperature [10, 11], due to its unique tilted energy-band structure [12-14] and generates DC current-induced anomalous Hall effect and nonreciprocal Hall effect behavior in thin-layered samples [15]. Based on its excellent electrical properties, it has shown great potential for applications in spintronic memories [16-19], quantum information devices [20, 21], and other fields. Recently, key progress has been made in the study of external-field modulated NLHE, in which Liu et al. successfully realized the multi-state modulation of third-order nonlinear Hall voltage by applying uniaxial strain, laying the foundation for the development of high-dimensional strain sensor devices [22]. However, there is still a gap in the systematic research on the full electric field modulation of third-order NLHE, which limits its application in programmable quantum devices significantly. The reason for this is that the dynamic reconfiguration mechanism of the BCP tensor under the action of electric field and its competition with impurity scattering have not yet been clarified, especially the modulation law of temperature on the dominant mechanism still lacks experimental evidence.

In this study, by constructing TaIrTe$_4$ single-crystal micro-nano devices, combining variable-temperature electric transport measurements (4 - 300 K) with scaling law analysis, we have revealed for the first time the regulation mechanism of the full electric field on the third-order NLHE. The main innovative discoveries as following: (1) temperature-induced sign reversal of the third-order NLHE response is observed, which confirms the competing dominant transition between BCP mechanism and scattering effect; (2) the quantitative correlation model of electric field strength (Edc) and response strength ( $E_\perp^{3\omega} / (E_\parallel^\omega)^3$ ) is established, which explains the modulation law of the electric field modulating the response strength at different temperatures; (3) a proposal of the modulation law of the response strength based on the BCP and impurity scattering effect, which provides a theoretical basis for the design of topological electronic devices. The results break through the technical limitations of the existing strain modulation and open up a new way for the development of novel electronically controlled quantum components.

**Results**

**Structure, electrical characteristics and Raman spectra for the TaTrTe$_4$ device.**

The atom structure of TaIrTe$_4$ exhibits a clear uniaxial mirror symmetry with mirror lines ($M_a$) along the lattice $b$-axis, as shown in Fig. 1a. The orientations of the crystals can be further verified by high-angle annular dark-field sanning transmission microscopy (STEM) image (see Fig. 1b). The STEM image imaged along the $b$-axis shows a mirror asymmetric structure, indicating that the crystal structure of the TaIrTe$_4$ is consistent with the bulk crystal structure. Figure 1c shows an optical microscope image of the device A1, which consists of mechanically exfoliated TaIrTe$_4$ flakes transferred onto a disk-shaped electrode with 12 contact pads.. The thickness of the flake is about 150 nm as measured by atomic force microscopy (AFM). The polarized Raman spectrum (see Supplementary Fig. S1) exhibits that the intensity of the Raman peak at 147 cm$^{-1}$ can be used for determining the $a$-axis and $b$-axis of the TaIrTe$_4$ crystals [17, 23].

The schematic diagram of device measurement is shown in Fig. 1d. Based on the transverse and longitudinal first harmonics (see Supplementary Fig. S2), the transverse and longitudinal resistance anisotropy of the device is shown in Fig. 1e. We fit the equations relating the resistance to the crystal point group $Pm$: $R_{\parallel}(\theta) = R_b \cos^2 \theta + R_a \sin^2 \theta$ and $R_{\perp}(\theta) = (R_a - R_b) \sin \theta \cos \theta$, where $R_a$ and $R_b$ represent the resistance along the lattice $a$-axis and $b$-axis, respectively. The good fitting results show that the crystal axial direction is consistent with the existing reports [11]. Figure 1f shows the third-order and second-order nonlinear anomalous Hall voltage signals at $\boldsymbol{E^{\omega}}$ $\parallel \theta = 30°$ and $E_{dc} = 0$. The second-order NLHE signals show relatively weak, while the third-order NLHE signals are very significant, which is consistent with previous reports [1, 6, 7, 11].

**Third-order NLHE of the TaTrTe₄ device.**

First, we measured the third-order nonlinear Hall voltage signal of the device A without applying a DC electric field $\boldsymbol{E}_{\text{dc}}$. To confirm the authenticity of the detected third-order NLHE signal, we ruled out factors that could potentially cause the third-order signal, such as capacitive effects, thermal effects, and capacitive coupling effects (see Supplementary Fig. S3). Figure 2a shows the curve of the third-order nonlinear Hall signal $V_{\perp}^{3\omega}$ as a function of $I^{\omega}$ at approximately room temperature (290 K). The results indicate that there is a linear relationship between $V_{\perp}^{3\omega}$ and $(I^{\omega})^3$ (Fig. 2b), further confirming that the measured signal is the intrinsic third-order nonlinear Hall voltage signal of TaIrTe₄. Note that when $I^{\omega}$ is applied along the crystal's *a*-axis or *b*-axis, the measured $V_{\perp}^{3\omega}$ is very weak, almost disappearing.

The angle dependence of the third-order nonlinear Hall response strength $\frac{V_{\perp}^{3\omega}}{V_{\parallel}^3}$ can be fitted using a formula derived based on the *Pmn2₁* symmetry [1] :

$$\frac{V_{\perp}^{3\omega}}{V_{\parallel}^3} \propto \frac{\cos\theta\sin\theta\left[(3\chi_{21}r^2-\chi_{11})\cos^2\theta+(\chi_{22}r^4-3\chi_{21}r^2)\sin^2\theta\right]}{(\cos^2\theta+r\sin^2\theta)^3}. \tag{1}$$

Here, $\chi_{ij}$ are elements of the third-order susceptibility tensor, and $r=\rho_b/\rho_a$ is the ratio of the resistivity of the *b*-axis to the *a*-axis. The fitting results are shown in Supplementary Fig. S4. The significant lattice-dependent characteristics indicate that the third-order nonlinear Hall effect is an intrinsic property of TaIrTe₄ crystals.

Subsequently, we investigated the temperature dependence of the third-order NLHE. As the temperature decreased, we observed a reversal in the sign of $V_{\perp}^{3\omega}$ (Fig. 2c). By studying the temperature dependence of the third-order nonlinear Hall response strength $V_{\perp}^{3\omega} / (V_{\parallel}^{\omega})^3$, we determined that the critical temperature is approximately 23

K (Fig. 2d). This result differs from the recent report by Wang et al. on the increase in the third-order nonlinear Hall signal with decreasing temperature in $T_d$-TaIrTe$_4$ [11], which may be related to the competition between intrinsic BCP and scattering effects.

**Electric field control of third-order NLHE.**

We investigated the control of the electric field on the third-order NLHE. Figure 3a shows the relationship curves between $V_\perp^{3\omega}$ and $I^\omega$ under different $E_{dc}$ conditions at 4 K. All $V_\perp^{3\omega}$ and $(I^\omega)^3$ exhibit linear relationships, as shown in Fig. 3b. This indicates that the electric field has a significant effect on the modulation of the third-order nonlinear Hall signal, with the measured signal gradually weakening as the electric field increases. Figure 3c shows the relationship between the third-order nonlinear Hall response strength $V_\perp^{3\omega} / (V_\parallel^\omega)^3$ and temperature and applied electric field. As can be seen from Fig. 3c, as the applied electric field increases, the critical temperature corresponding to the sign change of the third-order nonlinear Hall signal (where $V_\perp^{3\omega} / (V_\parallel^\omega)^3 = 0$) gradually decreases. Figure 3d shows the relationship between the applied electric field of different magnitudes and $V_\perp^{3\omega} / (V_\parallel^\omega)^3$ at 4 K, indicating that the third-order nonlinear Hall response strength and the applied electric field exhibit a linear relationship.

We also investigated the modulation of the third-order NLHE by applying external electric fields in different directions. Figure 4a shows that, at $E_{dc}$ = 0.3 kV/cm, changing the direction angle $\alpha$ of the applied external electric field results in a significant change in the magnitude of the $V_\perp^{3\omega}$ signal compared to $V_\perp^{3\omega}$ signals without an external field. All $V_\perp^{3\omega}$ and $(I^\omega)^3$ satisfy a linear relationship (Fig. 4b). We plotted the dependence of the third-order nonlinear Hall response strength on the external electric field angle and temperature, as shown in Fig. 4c. Figure 4c shows that at $E_{dc}$ = 0.3 kV/cm, the third-order nonlinear Hall response strength varies with the external field angle at different temperatures, with a period of 180°. This electric field-controlled effect persists up to room temperature (see Supplementary Fig. S5). Figure 4d shows the variation of the third-order nonlinear Hall response strength with the applied electric field angle at 4 K,

clearly exhibiting the 180° angle-dependent characteristic. Subsequently, we calculated

the relative change in the electric field-modulated response strength $\lambda$, where $\lambda =$

$\frac{V_{\perp}^{3\omega}/(V_{\parallel}^{\omega})^3(E_{dc}=0) - V_{\perp}^{3\omega}/(V_{\parallel}^{\omega})^3(E_{dc})}{V_{\perp}^{3\omega}/(V_{\parallel}^{\omega})^3(E_{dc}=0)} \times 100\%$. Within one period, when $E_{dc}$ = 0.3 kV/cm and $E_{dc}$ is along the 30° direction, $\lambda$ reaches a maximum value of approximately 65.3%, indicating that the electric field has a significant control effect on third-order NLHE.

Similar results were obtained for the devices B with the thickness of about 70 nm (see

Supplementary Fig. S6), confirming the reliability of the results.

**Mechanism of electric-field control of third-order NLHE.**

To explore the mechanism of electric field-controlled third-order nonlinear Hall effect, we analyzed the temperature-dependent third-order NLHE. According to the scaling law, the nonlinear Hall response strength $E_{\perp}^{3\omega}/(E_{\parallel}^{\omega})^3$ and longitudinal

conductivity $\sigma$ satisfy the following relationship [1, 11]:

$$|E_{\perp}^{3\omega}|/(E_{\parallel}^{\omega})^3 = \xi\sigma^2 + \eta, \tag{2}$$

where $\xi$ and $\eta$ are constants. Figure 5a shows the trend of the normalized resistivity $\rho/\rho_o$

of the device as a function of temperature, with $\rho_o$ being the resistivity at 4 K. For $E_{dc}$ =

0, as the temperature decreases from near room temperature to low temperatures, $\rho/\rho_0$

gradually decreases, which is attributed to lattice scattering dominating the process. When the temperature drops below approximately 24 K, $\rho/\rho_0$ begins to decrease slowly, with impurity scattering becoming the dominant contribution, consistent with previous reports [10]. However, when $E_{dc} = 0.5$ kV/cm is applied, the temperature must drop to approximately 19 K before impurity scattering becomes the dominant mechanism.

Figure 5b shows the relationship between $|E_{\perp}^{3\omega}|/(E_{\parallel}^{\omega})^3|$ and $\sigma$. Since the sign of the third-order NLHE reverses as the temperature decreases, there are two regions: a high-temperature region where $E_{\perp}^{3\omega}/(E_{\parallel}^{\omega})^3 < 0$ and a low-temperature region where $E_{\perp}^{3\omega}/(E_{\parallel}^{\omega})^3 > 0$ (see Supplementary Fig. S7). Considering the influence of the critical temperature, we performed linear fits using equation (2) for the high-temperature region (above the critical temperature) and the low-temperature region (below the critical temperature), respectively, obtaining the parameters $\xi$ and $\eta$ for the high-temperature and low-temperature regions, respectively, as shown in Figs. 5c and 5d. In the high-temperature region, as $E_{dc}$ increases, $\xi$ decreases slightly, while $\eta$ remains unchanged. For the low-temperature region, as $E_{dc}$ increases, $\xi$ gradually increases, while $\eta$ gradually decreases.

The third-order nonlinear response coefficient $\chi^{3\omega}$ satisfies the relationship $|E_{\perp}^{3\omega}|/(E_{\parallel}^{\omega})^3 \approx \chi^{3\omega}/\sigma$ [1, 11]. Since $\sigma$ is linearly related to the scattering time $\tau$, according to equation (2), $\chi^{3\omega}$ has two contributions, namely $\tau$ and $\tau^3$. The third-order nonlinear response coefficient exhibits linear response relationships with both $\tau$ and $\tau^3$ [2, 11, 24], corresponding to BCP-like and Drude-like contributions, respectively. Therefore, the contribution to the third-order nonlinear response strength

can be divided into BCP-like and Drude-like components, corresponding to the coefficients $\eta$ and $\xi$, respectively [3, 11].

Figures 5e and 5f show the BCP-like and Drude-like contributions at $E_{dc} = 0$ and 0.5 kV/cm, respectively. Without an external field, as the temperature decreases, near the critical temperature, the dominant contribution transitions from BCP-like to Drude-like, leading to a sign reversal in the third-order NLHE and reflecting the enhanced impurity scattering effect below the critical temperature. This is consistent with the temperature dependence of the device's resistivity shown in Fig. 5a. After applying an electric field, the transition of the dominant contribution persists. The insets in Figs. 5e and 5f illustrate the magnitudes of these two contributions in the high-temperature region. Additionally, after applying an electric field, above the critical temperature, the BCP-like contribution remains dominant but its magnitude slightly decreases (see the inset of Fig. 5f), leading to a weakened response strength of the third-order NLHE under electric field control (see Fig. 3c).

Below the critical temperature, compared with the absence of an external field, the values of the electric field-induced BCP-like and Drude-like contributions both decrease rapidly (see Fig. 5f), leading to a significant weakening of the response strength of the third-order NLHE under electric field control (see Fig. 3d). Additionally, the critical temperature without an external electric field is approximately 23 K, while at $E_{dc} = 0.5$ kV/cm, the critical temperature decreases to 17 K, which is consistent with the temperature at which the material transitions to residual resistance after applying an electric field (see Fig. 5a).

**Conclusions**

We have observed a sign reversal phenomenon of temperature-induced third-order NLHE in the type-*II* Weyl semimetal TaIrTe₄. Through scaling law analysis, we found that the response strength of third-order NLHE depends on temperature, with the high-temperature region dominated by BCP and the low-temperature region dominated by the impurity scattering effect. Additionally, applying an electric field can effectively regulate the response strength of third-order NLHE. At high temperatures, the reduction in BCP caused by the electric field leads to a decrease in the third-order NLHE response strength. At low temperatures, the electric field can simultaneously weaken both BCP and impurity scattering effects, thereby exerting a more significant regulatory effect on the third-order NLHE response strength. The electric field-induced regulation of third-order NLHE in TaIrTe₄ significantly enhances its application potential, offering a promising candidate for the development of novel electronic devices.

**Experimental Section**

Sample Preparation: First, thin flakes were obtained from bulk single-crystal $T_d$-TaIrTe₄ (HQ Graphene) using mechanical exfoliation. Then, circular Ti (5 nm)/Au (20 nm) structures were fabricated on a Si/SiO₂ (285 nm) substrate using electron beam lithography and electron beam evaporation techniques. Finally, the $T_d$-TaIrTe₄ flakes were transferred to the electrode using a dry transfer method. To prevent oxidation, both the exfoliation and transfer of the flakes were performed in a glove box filled with N₂ (water and oxygen content <0.1 ppm).

Sample Characterization: The thickness of TaIrTe₄ thin films was measured using AFM

(Dimension ICON). The crystal structure of a TaIrTe$_4$ flake was characterized using a Raman spectroscopy instrument with a laser wavelength of 532 nm. High-resolution STEM imaging was performed using a JEOL-ARM200F microscope equipped with an ASCOR aberration corrector and cold-ffeld emission gun, while operating at 200 kV. DC and sinusoidal AC currents at a frequency of 13.7 Hz were applied using a Keithley 2450 and 6221 source meters, respectively, while the first, second, and third harmonic signals were measured using a lock-in amplifier (Stanford S830).

Acknowledgement

This work was supported by the National Key R&D Program of China (Nos. 2023YFA1406603 and 2022YFA1403602), the National Natural Science Foundation of China (Nos. 12374112, T2394473, 52001169, and 61874060), the Open Foundation from Anhui key laboratory of low-energy quantum materials and devices, and the Innovation Project of Jiangsu Province (No. KYCX22_0918).

Conflict of Interest

The authors have no conflicts to disclose.

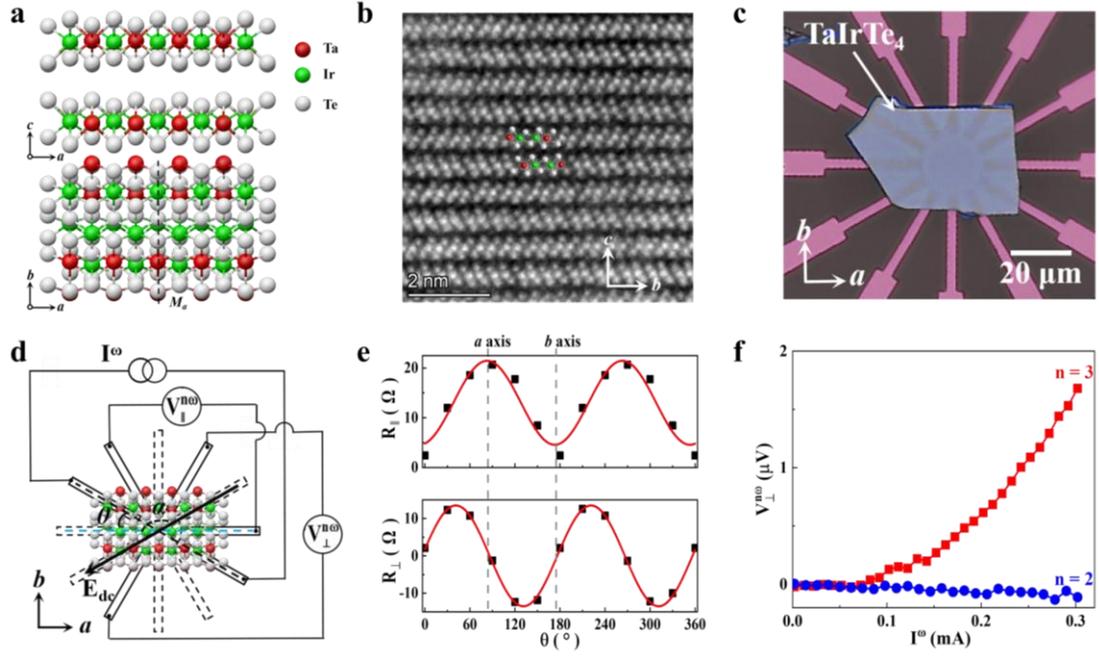

**Fig. 1**. **Structure, Scanning transmission electron micrograph and electrical characteristics for the TaTrTe₄ device. a**, Schematic of atomic structure of $T_d$-TaIrTe₄ viewed along the *a-c* plane and *a-b* plane, respectively. **b**, A high-angle annular dark-field cross-sectional scanning transmission electron micrograph of the TaIrTe₄ crystal along the *b-c* plane. The inset illustrates the overlap with the atomic configuration of TaIrTe₄. **c**, Optical microscope image of device A. **d**, Measurement configuration for resistances and second- and third-order NLHEs. $\theta$ is defined as the angle between $\boldsymbol{E}^{\omega}$ and the *a*-axis, and $\alpha$ is defined as the angle between the direction of $E_{dc}$ and the *a*-axis. **e**, Anisotropy plots of the Hall resistance (R⊥) and longitudinal resistance (R∥) of the device A, with the red solid line representing the fitted curve. **f**, Second and third harmonic Hall voltage signals measured without an external electric field and with $\boldsymbol{E}^{\omega}$ ∥ $\theta = 30°$.

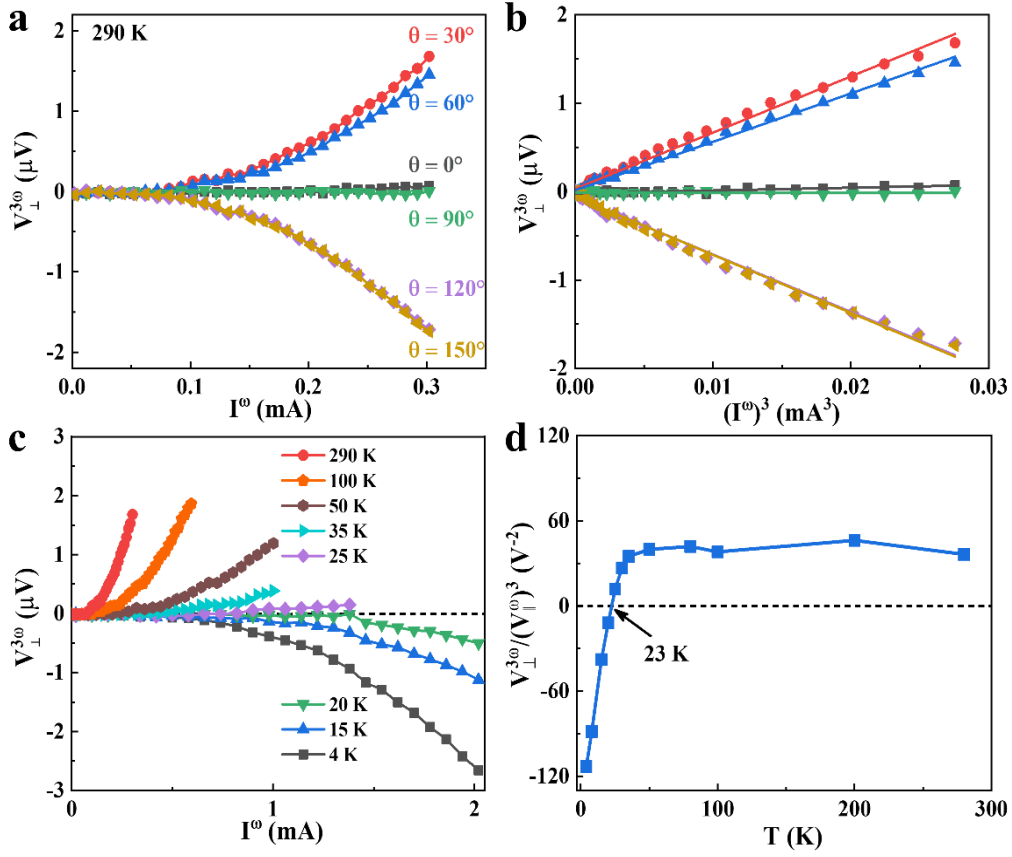

**Fig. 2**. **Third-order NLHE of the TaTrTe₄ device. a,** The dependence of the third-order Hall voltage $V_\perp^{3\omega}$ on $I^\omega$ at 290 K under $\boldsymbol{E}^\omega$ along different angles. **b,** $V_\perp^{3\omega}$ versus $(I^\omega)^3$ curves at 290 K under $\boldsymbol{E}^\omega$ along different angles. These lines represent the results of linear fitting. **c,** $V_\perp^{3\omega}$ versus $I^\omega$ curves at different temperatures under $\boldsymbol{E}^\omega$ along $\theta = 30°$. **d,** Temperature dependence of $V_\perp^{3\omega} / (V_\parallel^\omega)^3$.

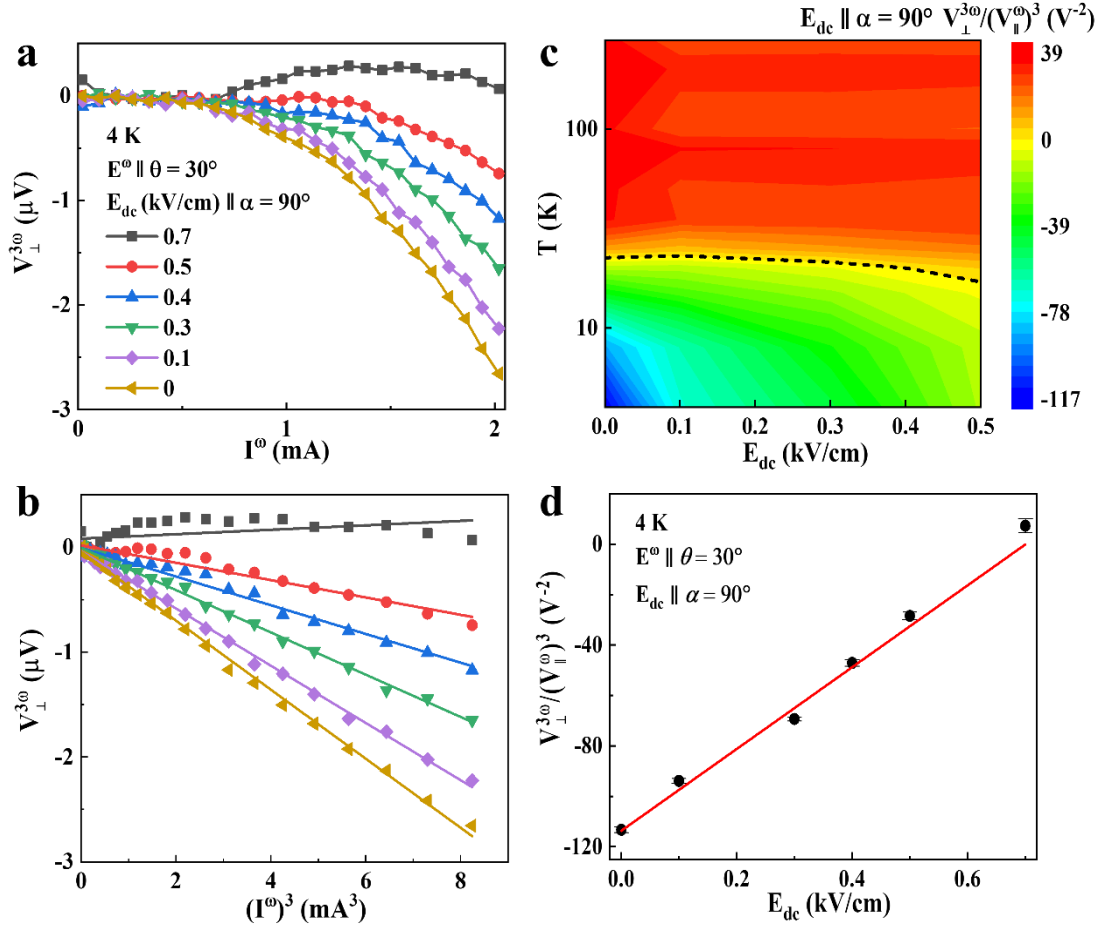

**Fig. 3**. **Electric field control of third-order NLHE. a,** The dependence of $V_\perp^{3\omega}$ on $I^\omega$ under different $\boldsymbol{E}_{dc}$ at 4 K with $\boldsymbol{E}^\omega \parallel \theta = 30°$ and $\boldsymbol{E}_{dc} \parallel \alpha = 90°$. **b,** The dependence of $V_\perp^{3\omega}$ on $(I^\omega)^3$ in **a**. **c,** The variation of $V_\perp^{3\omega} / (V_\parallel^\omega)^3$ with temperature (T) and $E_{dc}$ with $\boldsymbol{E}_{dc} \parallel \alpha = 90°$ and $\boldsymbol{E}^\omega \parallel \theta = 30°$. **d,** $V_\perp^{3\omega} / (V_\parallel^\omega)^3$ versus $\boldsymbol{E}_{dc}$ at 4 K with $\boldsymbol{E}^\omega \parallel \theta = 30°$ and $\boldsymbol{E}_{dc} \parallel \alpha = 90°$. The lines in **a** and **d** represent the results of linear fitting.

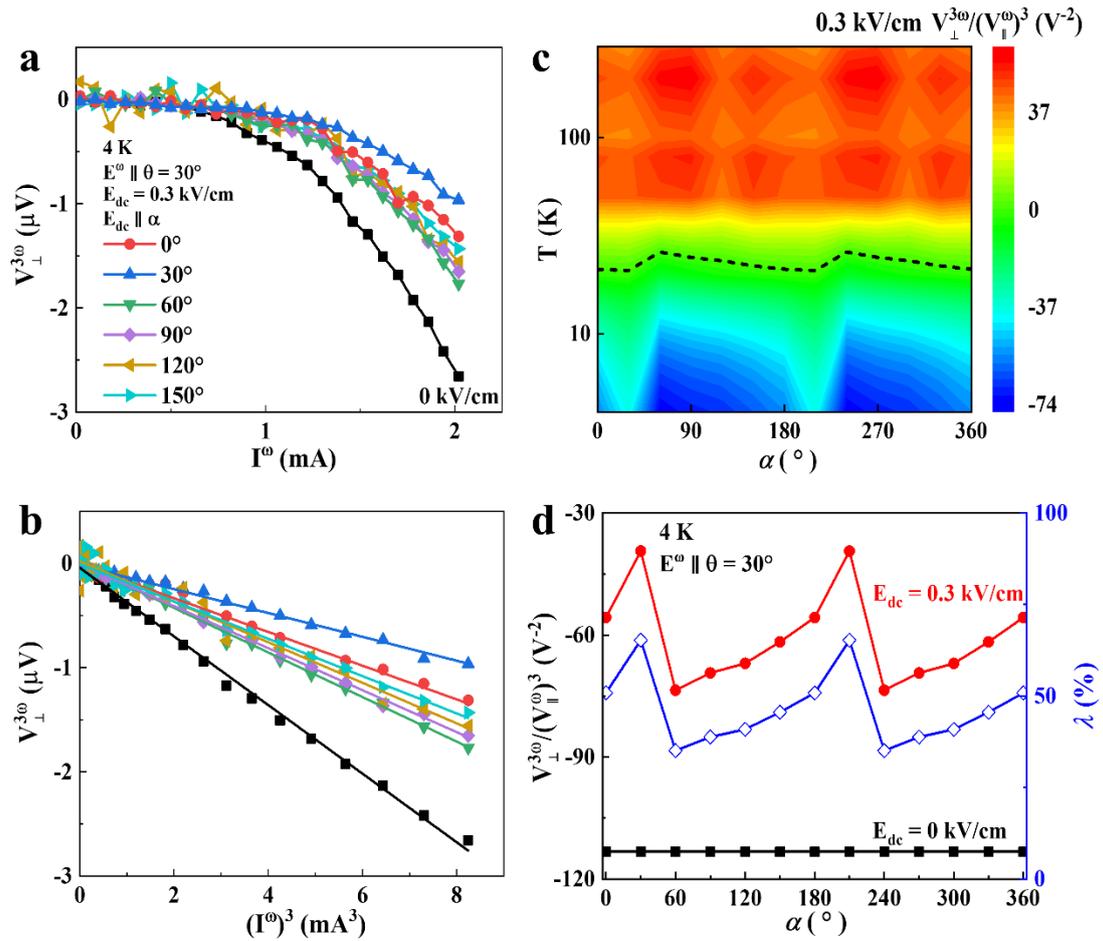

**Fig. 4**. **Angle-dependent characteristics of electric filed control of the third-order NLHE**. **a**, $V_{\perp}^{3\omega}$ versus $I^{\omega}$ under different external field directions at 4 K under $\boldsymbol{E}^{\omega} \parallel \theta = 30°$ and $E_{dc} = 0.3$ kV/cm. **b**, $V_{\perp}^{3\omega}$ versus $(I^{\omega})^3$ in **a**. **c**, The nonlinear Hall response $V_{\perp}^{3\omega} / (V_{\parallel}^{\omega})^3$ as a function of temperature and angle $\alpha$ at $E^{\omega} \parallel \theta = 30°$ and $E_{dc} = 0.3$ kV/cm. **d**, Third-order nonlinear Hall response $V_{\perp}^{3\omega}/\left(V_{\parallel}^{\omega}\right)^3$ versus $\alpha$ at 4 K and under $E^{\omega} \parallel \theta = 30°$ and $E_{dc} = 0, 0.3$ kV/cm. The blue data indicates the relative change in the electric field modulation parameter $\lambda$.

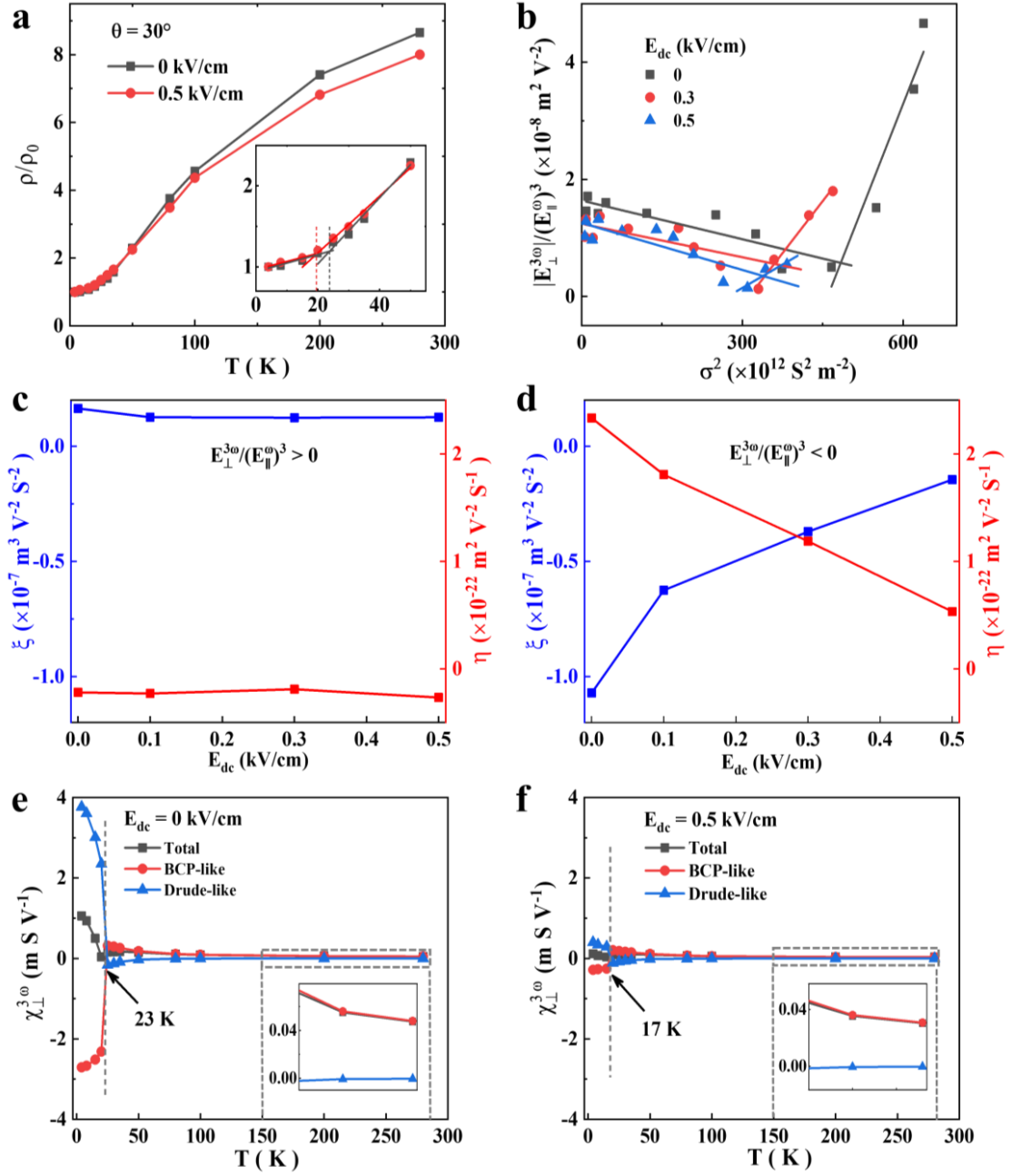

**Fig. 5. Scaling law analysis of the third-order NLHE controlled by an electric field.**

**a**, The temperature-dependent curves of the normalized resistivity $\rho/\rho_o$ with $\boldsymbol{E}^\omega \parallel \theta = 30°$, $E_{dc} \parallel \alpha = 90°$, and $E_{dc} = 0$, $0.5$ kV/cm. The inset is an enlarged view of the 4 - 50 K region, where the solid lines represent the results of linear fitting. **b**, $|E_\perp^{3\omega}|/(E_\parallel^\omega)^3$ as a function of $\sigma^2$ under $\boldsymbol{E}^\omega \parallel \theta = 30°$, $E_{dc} \parallel \alpha = 90$, and $E_{dc} = 0$, $0.3$ and $0.5$ kV/cm. The solid lines represent the results of linear fitting. **c, d**, The parameters $\xi$ and $\eta$ versus $E_{dc}$ for $E_\perp^{3\omega} / (E_\parallel^\omega)^3 > 0$ and $E_\perp^{3\omega} / (E_\parallel^\omega)^3 < 0$, respectively. **e**, **f**, The temperature dependence of the BCP-like, Drude-like, and total response coefficients under $E_{dc} = 0$ and 0.5 kV/cm, respectively. The gray vertical lines indicate the critical temperature points, and the insets are enlarged views of the gray-boxed regions.